# Трехмерные эффекты в динамике аномальных морских волн: численное исследование

В. П. Рубан[1]

Учреждение Российской академии наук
Институт теоретической физики им. Л. Д. Ландау РАН,
142432 Московская обл., г. Черноголовка, просп. акад. Семенова 1А

18 августа 2011 г.

**Аннотация**

Обсуждаются недавние результаты численного моделирования полностью нелинейных эволюционных уравнений для длинно-гребневых волн на глубокой воде, при котором наблюдалось самопроизвольное образование экстремально высоких волн --- т. н. волн-убийц. На нескольких примерах показано, что трехмерность движения жидкости весьма существенным образом влияет на процесс появления аномальных волн. В частности, при наличии узких и длинных волновых групп наиболее высокие волны-убийцы возникают в том случае, когда в начальном состоянии волновые фронты ориентированы под углом к направлению группы. ``Оптимальный'' угол ориентации гребней, приводящий к наибольшим волнам, зависит от начальной волновой амплитуды и от ширины группы, и составляет 18--28 градусов в практически важной области параметров. Кроме того, промоделирован механизм пространственно-временной фокусировки на фоне квазислучайного нелинейного волнового поля при разных значениях нелинейности.

PACS: 47.35.Bb, 92.10.-c, 02.60.Cb

---

[1] E-mail: ruban@itp.ac.ru

# 1 Введение

Проблема аномально высоких, пространственно локализованных волн (известных как ``волны-убийцы"), внезапно возникающих на умеренно взволнованной поверхности океана и затем ``бесследно" исчезающих, привлекает к себе внимание на протяжении уже ряда лет (см., например, [1-5], где обсуждаются различные физические механизмы этого явления, а также приведены ссылки на многие имеющиеся по данной теме работы; подробные исследования отдельных вопросов содержатся также, например, в статьях [6-20]). Наиболее вероятно выглядящий (хотя и несколько упрощенный) сценарий данного явления предполагает, что на начальной его стадии основную роль играют линейные механизмы усиления поверхностных волн, такие как их взаимодействие с неоднородным течением либо обусловленная дисперсией пространственно-временная фокусировка. Эти механизмы вызывают предварительное увеличение волновой амплитуды на некотором участке морской поверхности, и таким образом создают достаточно высокую и широкую группу волн, которая, условно говоря, становится неустойчивой по отношению к т. н. модуляционной неустойчивости [21-23]. В неустойчивой группе запускается нелинейный динамический процесс, еще более концентрирующий механическую энергию, в результате которого и возникает отдельная аномальная волна. Таким образом, волна-убийца представляет собой конечную стадию развития модуляционной неустойчивости, что было подтверждено прямым численным моделированием точных уравнений движения для потенциальных двумерных (2D) течений идеальной жидкости со свободной поверхностью [24-27].

Однако многие трехмерные (3D) стороны явления остаются далекими от полного понимания. Частично трудности в теории трехмерных волн-убийц связаны с ``дефицитом" численных результатов, вызванным отсутствием компактных и явных уравнений движения двумерной свободной поверхности в трехмерном пространстве, которые были бы удобны для численного счета. Отсутствие явных и точных эволюционных уравнений приводит к довольно громоздким и ``медленным" реализациям имеющихся численных методов, основанных непосредственно на решении интегрального уравнения, связывающего поверхностное распределение потенциала скорости и нормальную к поверхности компоненту поля скорости. Хотя необходимо отметить недавние продвижения в этой области [28-34], но даже самые передовые из этих разработок не способны за приемлемое время рассчитать динамику водной поверхности в 2D области с линейными размерами порядка 40-50 длин волн на протяжении десятков и сотен волновых периодов, как это требуется для полноценного моделирования зарождения волн-убийц. Поэтому для аналитических и численных исследований трехмерной динамики гигантских волн был предложен ряд приближенных моделей. В частности, для слабо-нелинейных режимов широко используется нелинейное уравнение Шредингера (НУШ) [22] и различные его обобщения [16, 17, 35-37], которые все являются упрощениями более общего интегро-дифференциального уравнения Захарова, где учитываются перенормированные волновые процессы $2 \to 2$ (см. [22, 38, 39] и ссылки там). Но слабо-нелинейные модели очевидно не годятся для описания волн-убийц на последней стадии их развития. Вот почему несколько лет назад автором была разработана альтернативная, полностью нелинейная приближенная модель, основанная на ином малом геометрическом параметре, а именно на малости отклонения от плоского течения [40]. Другими словами, здесь предполагается узость углового распределения волнового спектра в горизонтальной Фурье-плоскости, в отличие от слабо-нелинейных теорий, предполагающих малую крутизну волн. Данная квази-двумерная модель пригодна для описания произвольно крутых длинно-гребневых волн, распространяющихся близко к направлению оси $x$ в горизонтальной координатной плоскости $(x, q)$ (символ $y$ будет нами использоваться для обозначения вертикальной координаты). Соответствующий численный метод имеет приемлемую скорость счета на современных персональных компьютерах [41], и с его помощью оказалось возможным исследовать такие интересные вопросы как появление ``дышащих'' волн-убийц в случайном волновом поле [42] (см. также пример на Рис.1), нелинейную стадию модуляционной неустойчивости, сопровождающуюся образованием специфических зигзагообразных когерентных структур, производящих экстремальные волны при взаимодействиях между собой [43] (см. также пример на Рис.2), а также два различных типа волн-убийц в т.н. слабо-скрещенных состояниях морской поверхности [44]. Дополнительные численные примеры имеются в недавней статье [45].

В настоящей работе автор продолжает исследование трехмерных эффектов в динамике экстремальных морских волн. Одной из основных наших целей будет показать на нескольких численных примерах, что даже в отсутствие явной геометрической фокусировки трехмерность движения жидкости способна кардинальным образом усилить вероятность появления волн-убийц. Для этого в разделе 2 мы исследуем упрощенную ситуацию, которая в терминах уравнения Захарова [22, 38, 39], описывающего комплексные амплитуды волнового поля $b(\kappa\mathbf{i} + \mu\mathbf{j}, t)$ [где $\mathbf{i}$ и $\mathbf{j}$ --- базисные векторы некоторой горизонтальной системы координат $(\xi_1, \xi_2)$, повернутой по отношению к ранее упоминавшейся координатной системе $(x, q)$],

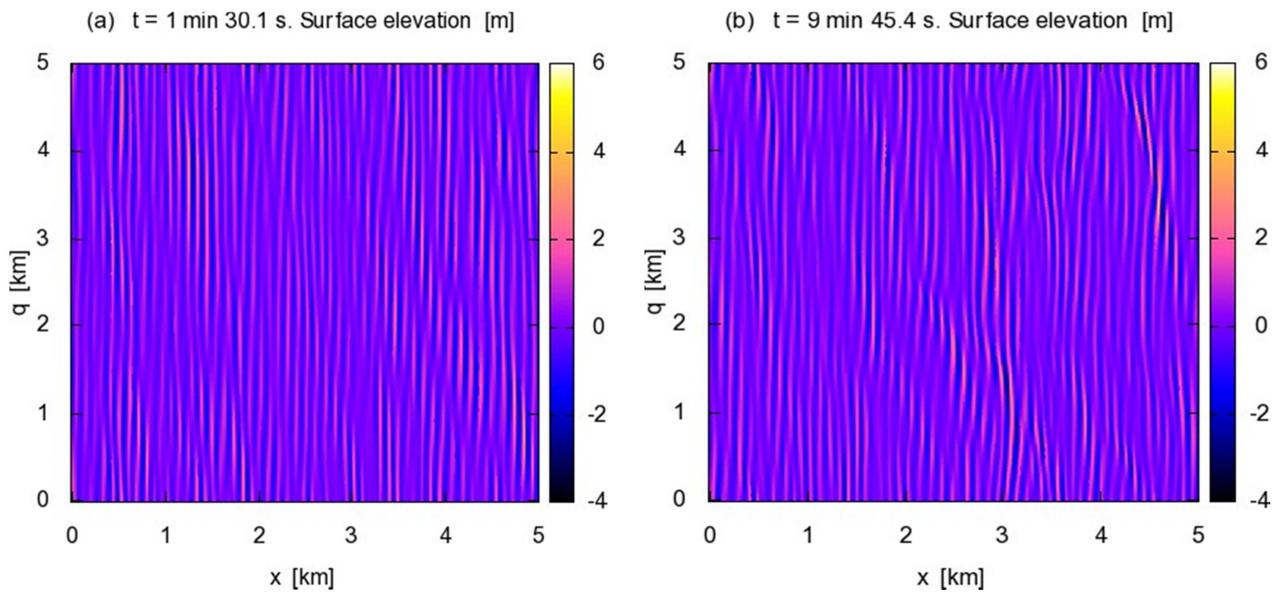

Рис. 1. Формирование ``дышащей'' волны-убийцы в квазислучайном волновом поле: a) нет аномально высоких волн; b) возникла волна-убийца высотой более 6 м при $x \approx 4.6$ км, $q \approx 3.5$ км. В данном численном эксперименте (немного отличающемся от описанного в работе [42]) большая волна просуществовала в осциллирующем режиме на протяжении около 25 волновых периодов, сделав за это время 12 колебаний.

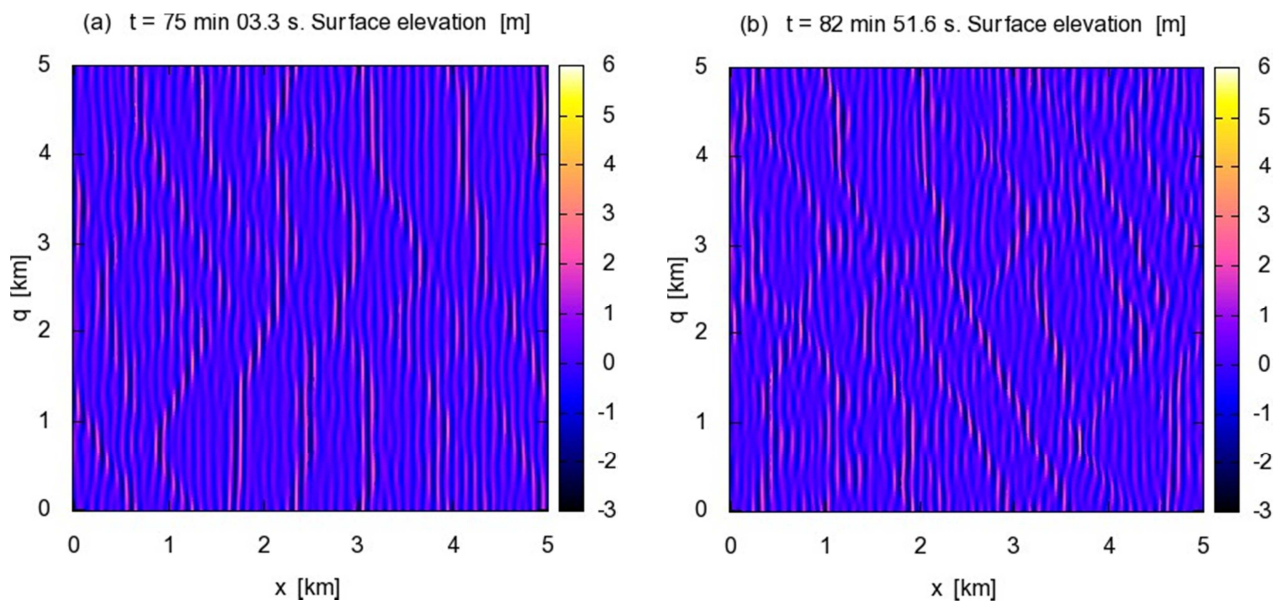

Рис. 2. Зигзагообразные структуры, возникающие на нелинейной стадии модуляционной неустойчивости: a) нет больших волн; b) аномальная волна высотой около 5.4 м сформировалась на ``повороте'' зигзага при $x \approx 3.7$ км, $q \approx 0.8$ км.

соответствует решениям вида $b(\kappa\mathbf{i}+\mu\mathbf{j},t) = \delta(\mu+m)\beta_\kappa(t)$, с постоянной $m$ ($\delta$ есть функция Дирака; очевидно, такая редукция допускается уравнением Захарова, учитывающим только процессы $2 \to 2$). В результате получается одномерная гамильтонова динамическая система

$$i\dot{\beta}_\kappa = \left[g^2(\kappa^2+m^2)\right]^{1/4}\beta_\kappa + \int \widetilde{T}^{\kappa,\kappa_1}_{\kappa_2,\kappa_3}\delta(\kappa+\kappa_1-\kappa_2-\kappa_3)\beta^*_{\kappa_1}\beta_{\kappa_2}\beta_{\kappa_3}d\kappa_1 d\kappa_2 d\kappa_3, \quad (1)$$

где $g$ --- ускорение свободного падения, а ``одномерная'' вершина 4-волнового взаимодействия $\widetilde{T}^{\kappa,\kappa_1}_{\kappa_2,\kappa_3}$ представляет собой соответствующую редукцию ``двумерной'' вершины $\hat{T}(\mathbf{k},\mathbf{k}_1;\mathbf{k}_2,\mathbf{k}_3)$ уравнения Захарова [22, 38, 39]:

$$\widetilde{T}^{\kappa,\kappa_1}_{\kappa_2,\kappa_3} = \hat{T}(\kappa\mathbf{i}-m\mathbf{j},\kappa_1\mathbf{i}-m\mathbf{j};\kappa_2\mathbf{i}-m\mathbf{j},\kappa_3\mathbf{i}-m\mathbf{j}). \quad (2)$$

Мы видим, что эффективный одномерный закон дисперсии волн в этом случае есть

$$\Omega_\kappa = \left[g^2(\kappa^2+m^2)\right]^{1/4}. \quad (3)$$

Важно отметить наличие (двух) точек перегиба на данной дисперсионной кривой, что приводит к нетривиальным эффектам в динамике нелинейных волн, о чем далее и пойдет речь.

Если мы стартуем с узкого спектрального распределения комплексного поля $\beta_\kappa$, сосредоточенного вблизи $\kappa_0 = [k^2-m^2]^{1/2}$, так что длина волны есть $\lambda = 2\pi/k$, то первая волновая гармоника определяется обратным Фурье-преобразованием вида

$$\int \beta(\kappa,t)\sqrt{(2\Omega_\kappa/g)}\exp(i\kappa\xi_1)d\kappa = B(\xi_1-V_{gr}t,t)\exp(-i\omega t + i\kappa_0\xi_1), \quad (4)$$

причем $\omega = \sqrt{gk}$ и $V_{gr} = d\Omega/d\kappa_0$. Волновая огибающая $B(x_1,t)$ на начальном этапе эволюции может быть приближенно описана нелинейным уравнением Шредингера (см. [22])

$$\frac{i}{\omega}\frac{\partial B}{\partial t} = \frac{[\cos^2\theta - 2\sin^2\theta]}{8k^2}\frac{\partial^2 B}{\partial x_1^2} + \frac{k^2}{2}|B|^2 B, \quad (5)$$

где угол $\theta = \arcsin(m/k)$ определяет ориентацию волновых фронтов по отношению к координатной оси $\xi_2$ в начальный момент времени. При положительных значениях множителя $D(\theta) = [\cos^2\theta - 2\sin^2\theta]$ мы имеем фокусирующее НУШ, допускающее решения бризерного типа, которые примерно и соответствуют интересующим нас волнам-убийцам. Если мы рассматриваем волновую группу с начальной амплитудой $A$ и с шириной $N\lambda$, так что характерная крутизна волн есть $s = kA$, то условие возникновения волны-убийцы в такой группе имеет вид

$$sN/\sqrt{D(\theta)} \gtrsim 1. \quad (6)$$

При первом взгляде на это условие возникает мысль, что взяв угол $\theta$ вблизи критического значения $\theta_* = \arctan(1/\sqrt{2}) \approx 35.3^o$, при котором $D$ обращается в ноль, можно добиться подъема большой волны в произвольно узкой группе волн, например при $N=1\cdots2$. Это, разумеется, не так, поскольку узкие группы нельзя описывать уравнением (5), а должно применяться более точное уравнение (1), да и то лишь при условии достаточной малости волновой амплитуды, пока нелинейности старших порядков не вступили в игру. Тем не менее, как показывают обсуждаемые далее численные эксперименты [46], в практически важной параметрической области $0.10 \lesssim s \lesssim 0.14$ и $0.7 \lesssim sN \lesssim 1.0$ условие (6) качественно верно отражает общую тенденцию к появлению высоких волн-убийц в косых структурах при $\theta = 18\cdots28^o$, тогда как при малых углах $\theta$ (порядка нескольких градусов) наблюдается лишь незначительный рост амплитуды. В указанной области параметров возникают волны-убийцы с высотой $Y_{max} \approx 0.06\lambda$, при которой начинается процесс их опрокидывания, тогда как отношение $Y_{max}/A = (2.5\cdots3)$. В то же время $Y_{min} \approx -0.04\lambda$, что объясняется асимметрией ``гребень-впадина'', свойственной гравитационным поверхностным волнам.

Что касается вопроса о поведении волн в косых полосах при $\theta$ вблизи $\theta_*$, то мы вернемся к нему в конце 2-го раздела. Не оставим мы без внимания и такой ``классический'' механизм формирования аномальных волн, как пространственно-временная фокусировка (сюда же мы относим и геометрическую фокусировку). Несколько соответствующих примеров будут представлены в разделе 3.

## 2 Усиленный подъем больших волн в косых группах

Содержание этого раздела соответствует недавней работе [46].

Один из наиболее важных вопросов в теории трехмерных волн-убийц связан с их появлением в некогерентных состояниях волнового поля, когда типичные волновые группы не слишком высокие и/или широкие, так что $sN \lesssim 1$. Однако в случаях, когда огибающие $B(x,q,t)$ волновых групп имеют длину много большую, чем их ширина (удлиненные волновые группы, как в слабо-скрещенных состояниях волнения [44]; см. также [47]), тогда дополнительный параметр --- угол $\theta$ --- вступает в игру. В данном разделе мы рассмотрим идеализированные, бесконечно длинные волновые группы, которые приближенно соответствуют уравнению (1), однако решать (численно) мы будем не его, а уравнения квази-двумерной, полностью нелинейной модели [40, 41]. К описанию соответствующих численных экспериментов мы сейчас и переходим.

Итак, мы имеем полосы вдоль оси $\xi_2$ в координатной системе $(\xi_1, \xi_2)$, которая повернута на некоторый угол $\gamma$ относительно $(x, q)$-системы, причем $\gamma$ слегка меньше $\theta$. Таким образом, начальные волновые фронты ориентированы не точно вдоль оси $q$, но повернуты под небольшим углом (в несколько градусов) по часовой стрелке, тогда как сама волновая полоса ориентирована под углом $\gamma$ против часовой стрелки. Такое расположение выбирается по той причине, что возникающие экстремальные волны всегда ориентированы более полого к полосе, чем начальные гребни (см. Рис.3, а также [43, 45]), и поэтому выбор $\gamma < \theta$ приводит к более близкой ориентации высоких гребней по отношению к оси $q$, как это требуется для применимости используемой квази-двумерной модели [40, 41].

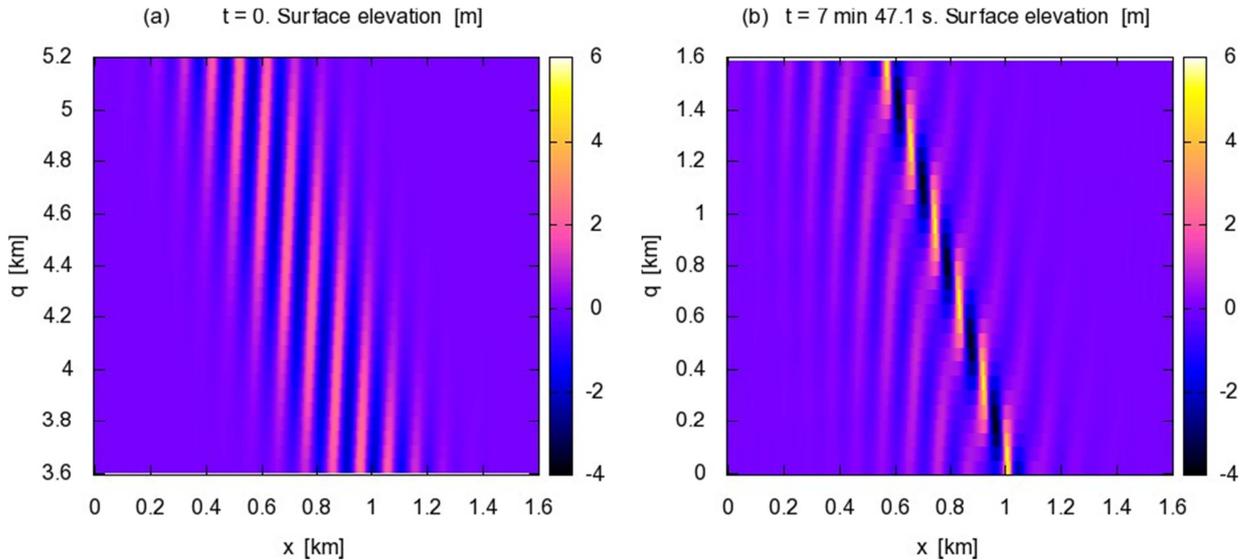

Рис. 3: Численный пример образования волн-убийц в косой волновой группе.

В начальном состоянии комплексная огибающая первой гармоники полагается чисто действительной и дается простым выражением

$$B(\xi_1, 0) \approx \frac{s}{k} \exp\left(\frac{-\xi_1^2}{2w^2 \lambda^2}\right). \qquad (7)$$

Параметр $N$ можно идентифицировать следующим образом: $N \approx 4w$. Мы также добавляем к начальному спектру малое двумерное возмущение со случайными фазами, подобно тому, как это делалось в [42, 43], чтобы быть уверенными, что наши результаты устойчивы по отношению к наличию шума в начальных условиях.

Для удобства графического представления, мы выбираем $\lambda = 100$ m, так что соответствующий период волны есть $T = \sqrt{2\pi\lambda/g} \approx 8$ с. Вычислительная область имеет прямоугольную форму $L_x \times L_q$, с периодическими граничными условиями по обоим координатам и $L_x = 2$ км. Для параметра $L_q$ брались несколько различных значений ($L_q = 4,5,6,7,8$ км), чтобы обеспечить квази-двумерный режим при различных углах $\theta$, по крайней мере на начальном этапе эволюции; кстати сказать, $\gamma = \arctan(L_x/L_q)$. Исключение составляет случай малых $\theta$, когда $L_q = L_x$, а $\gamma = 0$. Например, на Рис.3 показаны два участка полной области 2 км $\times$ 7 км.

Вычисления проводились на современных персональных компьютерах с использованием численного метода, описанного в работе [41]. Конечное разрешение достигало 16000×3000 точек в тех случаях, когда экстремальные волны приближались к моменту опрокидывания (начало опрокидывания характеризуется внезапным ростом максимальной волновой крутизны после достижения критического значения $s_* \approx 0.5$ рад, которое соответствует крутизне предельной волны Стокса). Некоторые из полученных численных результатов представлены на Рисунках 3-5. В частности, Рис.3(a) показывает карту свободной поверхности при $t = 0$, для $s = 0.14$, $N \approx 6$, и $\theta \approx 18.4^o$, тогда как Рис.3(b) представляет собой карту водной поверхности в более поздний момент времени, около нескольких десятков волновых периодов, когда экстремальные волны и глубокие впадины образовали косую структуру, напоминающую взятый в отдельности один из двух симметричных рядов волн после идущего корабля. Следует отметить, что фрагменты похожих волновых полос возникают самопроизвольно на нелинейной стадии модуляционной неустойчивости, где они образуют случайные зигзагообразные узоры (см. Рис.2), причем волны-убийцы возникают главным образом в углах зигзагов [43, 45].

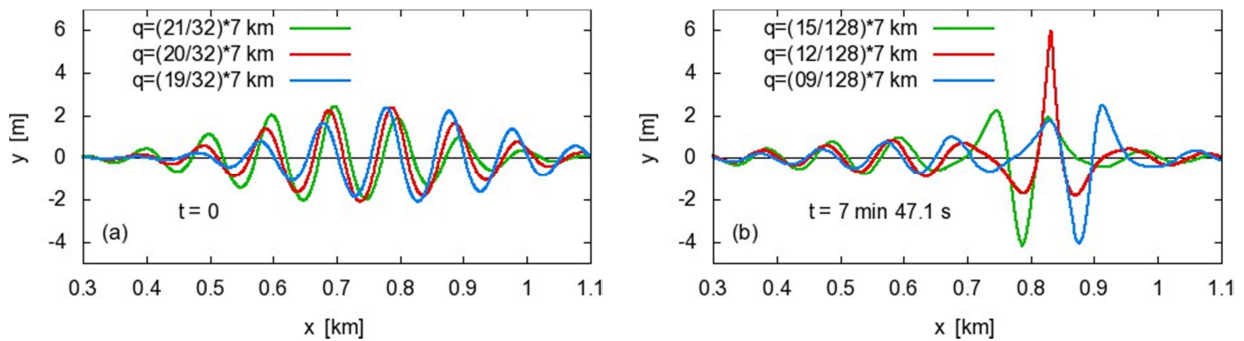

Рис 4: Некоторые волновые профили, соответствующие Рисунку 3. Волны-убийцы концентрируют энергию, и поэтому амплитуда остальных волн в группе понижается.

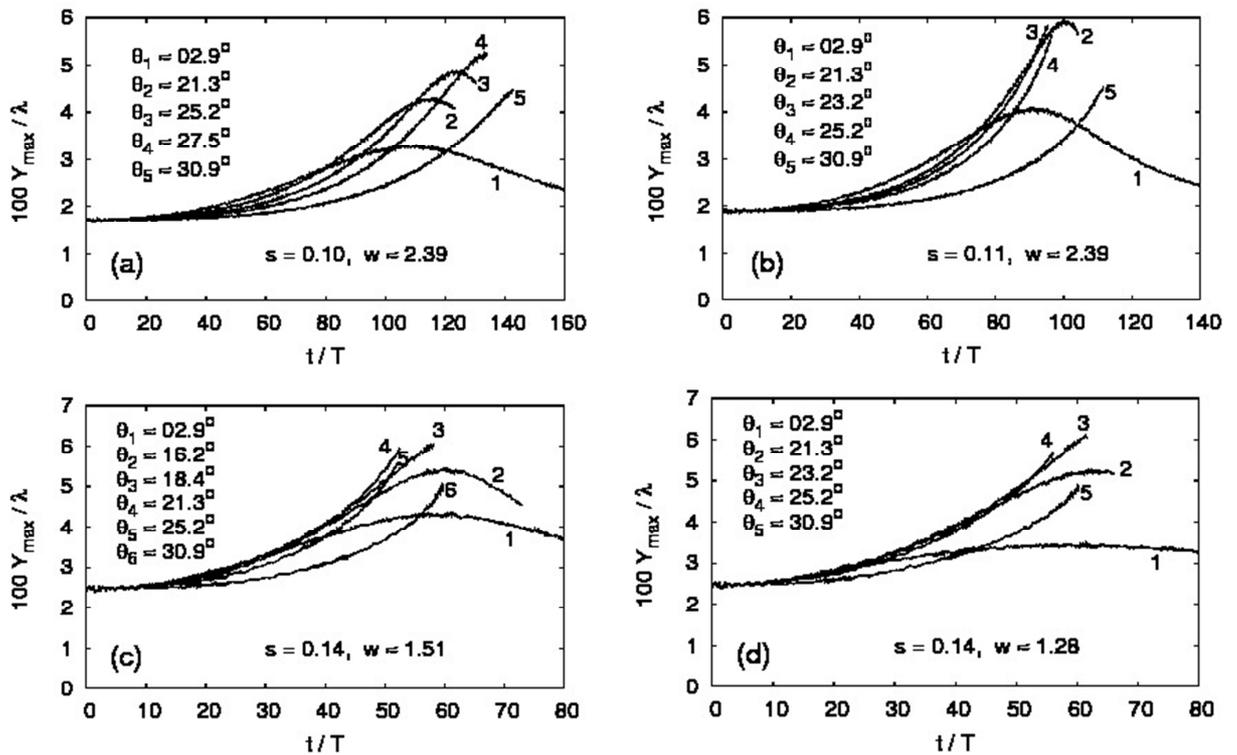

Рис 5: Максимальный подъем свободной поверхности в зависимости от времени в косых волновых полосах с заданными шириной и крутизной, для различных углов $\theta$. Волны, показанные на Рис.3(b) и Рис.4(b), соответствуют зависимости `3' на Рис.5(c), в точке, взятой немного раньше окончания кривой. Опрокидывание волн имеет место в конце экспериментов a5, b3, b4, b5, c3, c4, c5, c6, d3, d4 и d5.

На Рис.4 представлены некоторые волновые профили с Рис.3, которые подчеркивают такие поразительные черты волн-убийц, как их сильную локализацию и высокое относительное увеличение амплитуды. В целом, существенно нелинейная природа рассматриваемого явления подтверждается в этих численных экспериментах.

Очень важен и интересен Рис.5, где дано сравнение зависимостей максимального подъема поверхности от времени для различных $\theta$. Ясно видно, что для $s$ и $w$, удовлетворяющих соотношению $4sw \lesssim 1$, наиболее ``совершенные'' волны-убийцы с $Y_{max} \approx 0.06\lambda$ возникают в тех случаях, когда $\theta$ достаточно велик, около $18\cdots28^o$, тогда как при малых $\theta$ лишь умеренный подъем волн имеет место, с последующим спадом. Следует здесь отметить, что при малых $\theta$, в отсутствие опрокидывания, вслед за спадом наблюдался второй подъем на более поздних временах (не показано), но в данной работе мы не обсуждаем такое квази-рекуррентное поведение, обусловленное сложным взаимодействием квази-солитонов, составляющих волновую группу.

Оптимальный для формирования высоких волн угол $\tilde{\theta}(s,w)$ возрастает, когда $s$ или $w$ убывает [ср. Рис.5(a) с Рис.5(b), и Рис.5(c) с Рис.5(d)]. Однако существует уже упоминавшийся критический угол $\theta_*$, вблизи которого поведение волн в группе становится более сложным. Наши компьютерные вычисления показали, что фактически при $\theta$ вблизи $\theta_*$ наиболее высокие волны в полосе становятся существенно более короткими по сравнению с $\lambda$, и в конце происходит их опрокидывание при высотах $Y_{\max}$ заметно меньших, чем $0.06\lambda$. Необходимо также отметить, что при этом нарушается квази-двумерный динамический режим (направление гребней сильно меняется вдоль координаты $\xi_1$, что практически выглядит как их излом; не показано), и поэтому, строго говоря, эта параметрическая область не может быть аккуратно исследована с помощью квази-двумерной модели.

Итак, в данном разделе было показано, что косая ориентация волновых фронтов в удлиненных группах поверхностных волн способствует усилению процесса образования экстремальных волн. Этот существенно трехмерный эффект наиболее заметен, когда $sN \lesssim 1$. В следующем разделе мы рассмотрим другой трехмерный эффект, а именно взаимодействие между дисперсией и нелинейностью в процессе образования аномальной волны в квазислучайном волновом поле за счет механизма пространственно-временной фокусировки.

## 3 Пространственно-временная фокусировка

Хорошо известно, что высокие группы волн могут возникать случайным образом за счет пространственно-временной фокусировки, когда некоторые волновые моды оказываются ``в фазе'' в данном месте и в данное время. По-видимому, именно этот, практически линейный механизм, обусловленный двумерным законом дисперии $\omega(k)=(gk)^{1/2}$, отвечает за основную долю возникающих волн-убийц в некогерентных состояниях морского волнения с относительно широким спектром. Однако следует отдавать себе отчет, что нелинейность все же способна заметным образом повлиять на такой процесс линейной фокусировки.

Для примера мы промоделируем в данном разделе ситуацию, когда в начальный момент времени пространственный спектр волнения состоит из суммы двух частей. Одна часть содержит в себе основную долю энергии, но соответствующие 2D Фурье-гармоники имеют (псевдо-) случайные фазы и поэтому в физической плоскости соответствующие волны распределены в статистическом смысле равномерно. Эта часть спектра создает квазислучайный фон, ``маскирующий'' зарождение волны-убийцы. Другая, относительно малая часть начального спектра имеет фазовый множитель вида $\exp[iT^*\omega(k) - i(\mathbf{k}\cdot\mathbf{x}_0)]$, так что при отсутствии нелинейных взаимодействий в момент времени $T^*$ в точке плоскости $\mathbf{x}_0$ образовалась бы большая волна. Заметим, что фазы суммарного спектра в целом мало отличаются от случайных, и поэтому такая постановка численных экспериментов достаточно хорошо имитирует фокусировку, которая могла бы иметь место в реальном случайном волновом поле.

Как и прежде, характерную длину волны берем равной 100 м, но вычислительная область теперь имеет размеры 5×5 км. Выбрав в качестве ``точки отсчета'' некоторое начальное состояние с описанными выше свойствами и характерной амплитудой волн 2 м (при этом конкретная зависимость абсолютных значений спектральных компонент от волнового вектора оказывается не очень важна), мы умножаем затем соответствующий спектр на общий множитель $c$ из набора $\{0.4, 0.8, 1.0, 1.2\}$, тем самым задавая меру участия нелинейности в процессе образования аномальной волны в отдельно взятом эксперименте. Некоторые из результатов такого моделирования, для пяти различных значений параметра $T^*$, представлены на Рис. 6-8. Для удобства сравнения результатов, начальные фазы случайной части волнового поля также соответственно ``подкручивались'', так что в линейном режиме динамика для разных $T^*$ отличалась бы только сдвигом по времени.

Рис.6, во-первых, показывает, что при увеличении общей амплитуды $c$ наибольшая высота волн достигается на временах, заметно меньших, чем ``расчетное'' время линейной фокусировки $T^*$. Дальнейшая динамика становится дефокусирующей, поскольку гребни высоких волн за счет нелинейной поправки к фазовой скорости становятся выпуклыми по направлению вперед (см. Рис. 7), в результате чего энергия волны начинает рассеиваться еще до наступления момента $T^*$. Лишь для $c \lesssim 0.4$, при начальной амплитуде 100-метровых волн менее 0.8 м, процесс фокусировки происходит в почти линейном режиме. Во-вторых, при увеличении $c$ достигаемая максимальная высота волн-убийц не вполне пропорциональна начальной амплитуде, а наблюдается тенденция к насыщению на ``высоте опрокидывания'' (около 6 м при длине волн 100 м). Кроме того, при фиксированной амплитуде $c$ и существенном увеличении $T^*$ высота подъема ожидаемой аномальной волны уменьшается, а

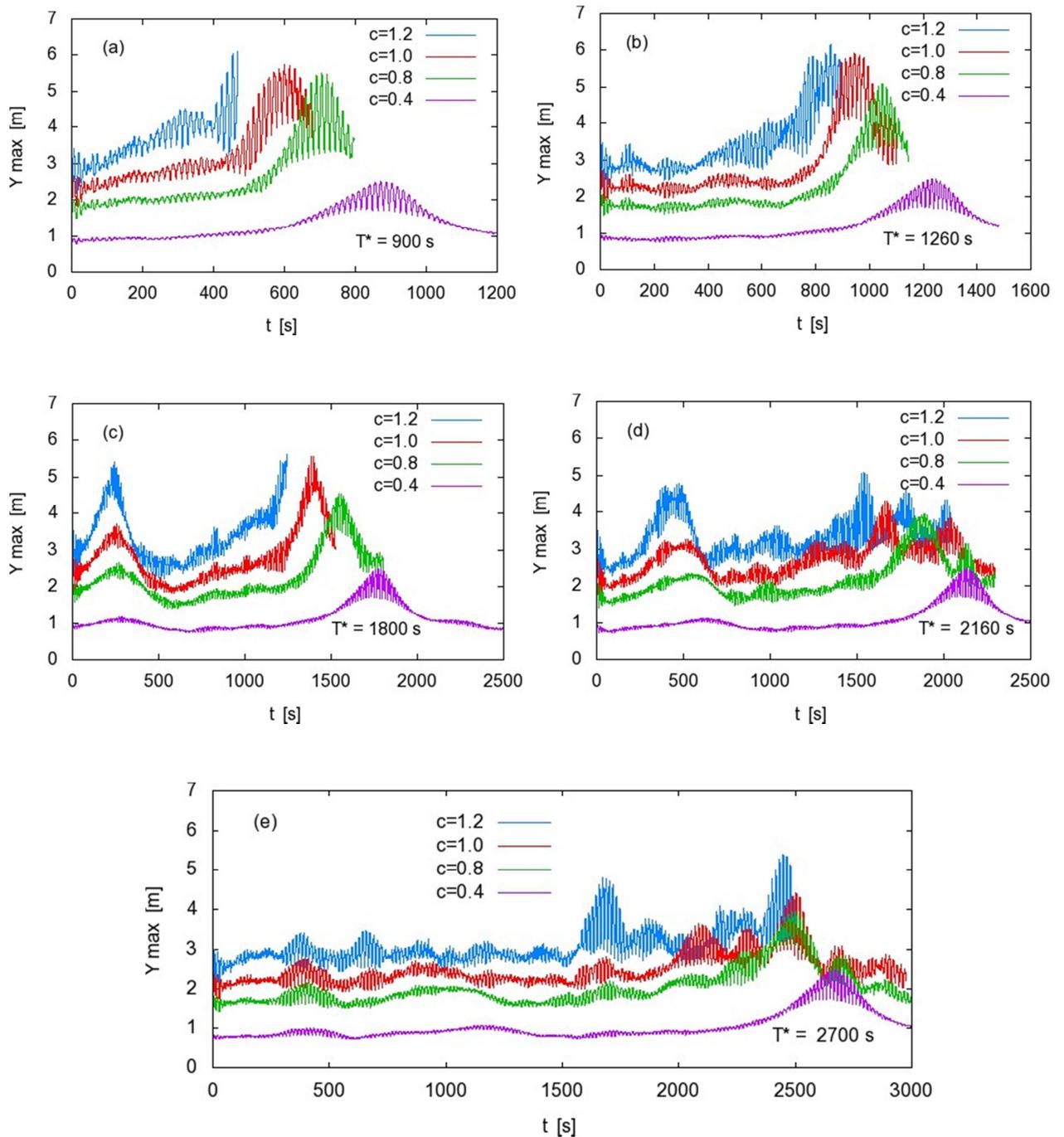

Рис. 6. Зависимости максимальной высоты волн от времени при различных значениях $c$ для пяти выборов $T^*$. Осцилляции кривых связаны с различием в два раза между фазовой и групповой скоростями. В экспериментах (c) и (d) дополнительно наблюдались достаточно высокие волны, не связанные с линейной фокусировкой. В экспериментах (d) и (e) при $c = 1.2$ и $c = 1.0$ вместо одной экстремальной волны в некогерентно сфокусированной группе возникали в разные моменты времени две-три более умеренных волны.

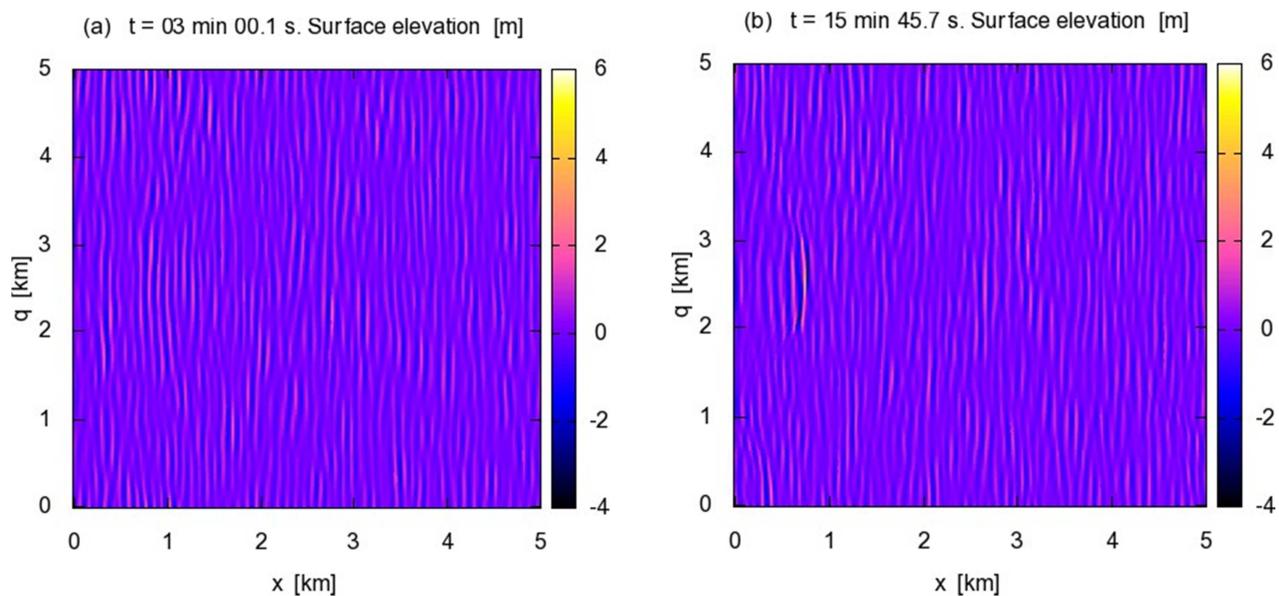

Рис 7. Численный пример образования аномальной волны путем фокусировки на случайном фоне волнения при $c = 1.0$, $T^* \approx 21$ мин: a) имеется группа умеренных волн с фокусирующей конфигурацией гребней в районе $x \approx 1$ км, $q \approx 2.5$ км; b) после прохождения пути в один период $L_x$, в группе образовалась отдельная волна-убийца высотой около 6 м при $x \approx 0.7$ км, $q \approx 2.5$ км.

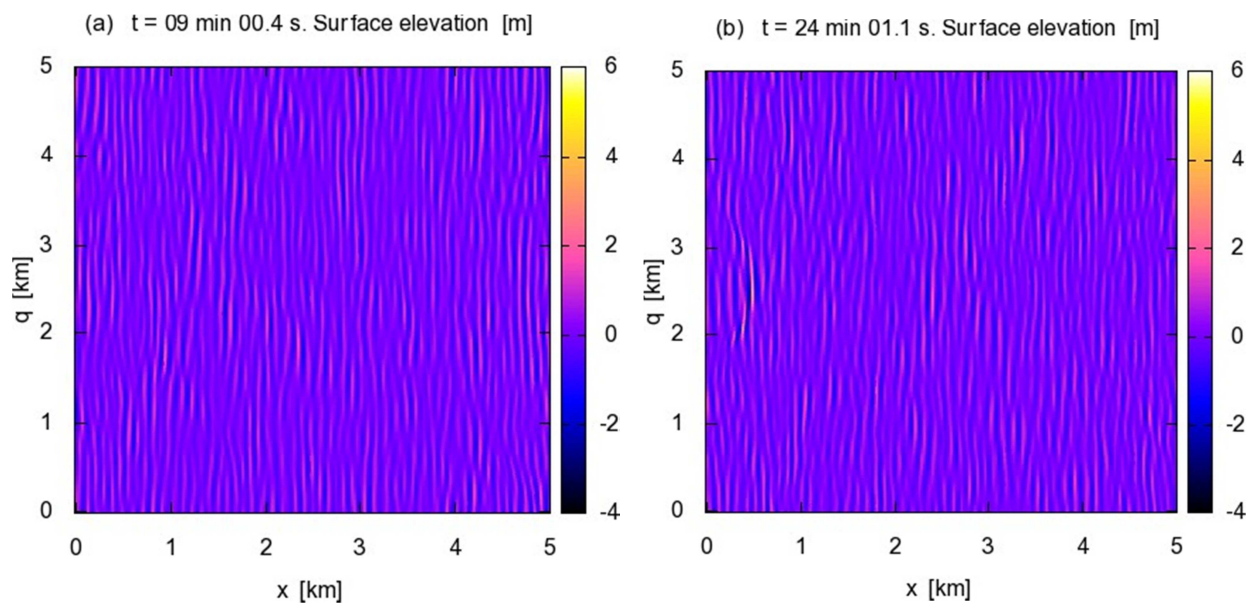

Рис 8. Фокусировка на случайном волновом фоне при $c = 1.0$, $T^* = 30$ мин. С увеличением времени $T^*$ сфокусированная волна искажается из-за более длительного нелинейного взаимодействия с фоном.

сама большая волна все более искажается (ср. Рис.7 и Рис.8), и даже вместо одной экстремальной волны в ``неаккуратно'' сфокусированной группе могут возникать поочередно несколько меньших волн, не вполне заслуживающих названия ``волны-убийцы'', но зато в других местах могут появляться ``непредвиденные'' аномальные волны. Указанные эффекты очевидно обусловлены нелинейными процессами. На Рис.7(a) показана свободная поверхность (для $c = 1.0$ и $T^* \approx 21$ мин) задолго до образования большой волны. Здесь ясно видна достаточно широкая группа не очень высоких волн, гребни которых имеют фокусирующую конфигурацию. На Рис.7(b) показана аномальная волна в момент наивысшего подъема (около 5.9 м), причем ее гребень уже сделался несколько выпуклым по направлению движения. На Рис.8 мы видим, что увеличение $T^*$ привело к искажению волны-убийцы, причем, что интересно, в ее ``портрете'' появились элементы, характерные для косых волновых структур, описанных в разделе 2.

Необходимо еще отметить, что для некоторых других начальных спектров, в которых при строго линейной динамике имел бы место подъем аномальной волны, в действительности нелинейность (относительно более сильная, чем на Рис.6-7) квазислучайного фона нарушала условие фазовой согласованности, в результате чего фокусировка ``сбивалась'', и волна-убийца в ее настоящем виде вообще не появлялась (не показано). По этой же причине искажается сфокусированная волна и уменьшается ее высота при увеличении $T^*$ в приведенных выше примерах.

## 4 Заключение

Приведенные в данной статье численные примеры свидетельствуют о том, что динамические процессы, приводящие к возникновению волн-убийц, в трехмерном пространстве оказываются значительно более сложным, чем можно было бы судить только по результатам исследований плоских течений. В частности, в узко-направленных волновых полях наличие поперечного горизонтального направления, вдоль которого нелинейность действует дефокусирующим образом, выделяет косо ориентированные полосы волновой огибающей как структуры, наиболее способствующие самопроизвольному образованию высоких волн. Ранее было отмечено, что такие косые структуры появляются на нелинейной стадии развития модуляционной неустойчивости, и кроме того, они возможны в т. н. слабо-скрещенных состояниях морской поверхности.

Численное моделирование возникновения 3D аномальных волн за счет пространственно-временной фокусировки на случайном фоне показывает, что такой фон дополнительным случайным нелинейным ``подкручиванием'' фаз воздействует на условие фазовой согласованности. Кроме того, нелинейность, с одной стороны, фокусирует энергию волн в группе вдоль направления движения, а с другой стороны, она искривляет фронты наиболее высоких волн, меняя тем самым характер геометрической фокусировки. Как следствие всех этих процессов, изменяется место и время появления гигантской волны, либо такая волна может вовсе не возникнуть при, казалось бы, благоприятных для нее начальных условиях.

По всей видимости, трехмерные эффекты в динамике аномальных морских волн далеко не исчерпываются приведеными выше примерами, тем более что в рамках нашей квази-двумерной модели могли быть исследованы только волновые поля с узко-направленными спектрами. В ``широко-угольных'' спектрах, да еще с несколькими максимумами, могут действовать и другие, пока неизученные механизмы образования гигантских волн. В данной области имеется большое количество нерешенных вопросов. В частности, до сих пор не выяснено, какие условия требуются для появления ``долгоживущих'' волн-убийц вроде той, что показана на Рис.1.

## Литература